\documentclass[preprint,12pt]{elsarticle}
\usepackage{amssymb}
\usepackage{amsmath}
\journal{JQSRT}

\newcommand{\cm}{cm$^{-1}$}
\newcommand{\um}{$\mu$m}

\begin{document}

\begin{frontmatter}

\title{Cross-sections for heavy atmospheres: H$_2$O continuum}

\author[inst1]{Lara O. Anisman}
\author[inst2]{Katy L. Chubb}
\author[inst3]{Jonathan Elsey}

\author[inst1]{Ahmed Al-Refaie}
\author[inst1]{Quentin Changeat}
\author[inst1]{Sergei N. Yurchenko}
\author[inst1]{Jonathan Tennyson}
\author[inst1]{Giovanna Tinetti}

\affiliation[inst1]{organization={Department of Physics and Astronomy, University College London},
            addressline={Gower Street}, 
            city={London},
            postcode = {WC1E 6BT},
            country={United Kingdom}}

\affiliation[inst2]{organization={Centre for Exoplanet Science, University of St Andrews},
            addressline={North Haugh}, 
            city={St Andrews},
            postcode={KY16 9SS},
            country={United Kingdom}}
            
\affiliation[inst3]{organization={Department of Meteorology, University of Reading},
            addressline={Earley Gate}, 
            city={Reading},
            postcode={RG6 6BB},
            country={United Kingdom}}

\begin{abstract}
 Most of the exoplanets detected up to now transit in front of their host stars, allowing for the generation of transmission spectra; the study of exoplanet atmospheres relies heavily upon accurate analysis of these spectra. Recent discoveries mean that the study of atmospheric signals from low-mass, temperate worlds are becoming increasingly common. The observed transit depth in these planets  is small and more difficult to analyze. Analysis of simulated transmission spectra for two small, temperate planets (GJ 1214 b and K2-18 b) is  presented, giving evidence for significant differences in simulated transit depth when the water vapor continuum is accounted for when compared to models omitting it. These models use cross-sections from the CAVIAR lab experiment for the  water self-continuum  up to 10,000 cm$^{-1}$; these cross-sections exhibit an inverse relationship with temperature, hence lower-temperature atmospheres are the most significantly impacted. Including the water continuum strongly affects transit depths, increasing values by up to 60 ppm, with the differences for both planets being detectable with the future space missions Ariel and JWST. It is imperative that models of exoplanet spectra move toward adaptive cross-sections, increasingly optimized for H$_2$O-rich atmospheres. This necessitates including absorption contribution from the water vapor continuum into atmospheric simulations.

\end{abstract}

\begin{keyword}
exoplanets \sep atmospheres \sep water vapor \sep opacities \sep continuum absorption \sep super-Earths \sep mini-Neptunes

\end{keyword}

\end{frontmatter}
\vspace{8mm}
\section{Introduction}

 In the preceding decade a variety of dedicated ground-based facilities and space missions, including CNES/ESA CoRoT, NASA Kepler and TESS, have employed the most prolific method for exoplanet detection, the \textit{transit method}, to detect upwards of 3000 exoplanets. As a planet transits in front of its host the starlight is modified, allowing for the generation of transmission spectra; the study of exoplanet atmospheres relies heavily upon accurate analysis of these. The influx of small planets discovered
 means that atmospheric signals of low-mass, temperate worlds are increasingly common. The signal acquired in transmission, known as the \textit{transit depth}, is smaller for these planets and as such more difficult to analyse.

 As the field of exoplanet atmospheric spectroscopy moves from the study of hot gaseous giants towards dissecting the difficult atmospheres of small ($\leq 10 $ $M_{\bigoplus}$), rocky and temperate ($\leq 500$ K) worlds, an increasingly nuanced approach is required. Our ability to accurately characterise the atmospheric constituents of these planets is dictated in a large part by the scale height of the atmosphere in question. Small planets, with cooler temperatures generally exhibit flatter spectra, for example GJ 1214\,b \cite{Kreidberg_2014}, and GJ 1132\,b \cite{LR_2020, Swain_2021}. On top of this, heavier atmospheres, with high mean-molecular-weight, for example water-based atmospheres, will have further suppressed scale heights. 
To complicate matters further, the presence of a thick cloud layer can veil a large portion of the atmosphere, rendering it opaque above a certain pressure. As a consequence, H$_2$-rich atmospheres with trace amounts of heavy molecules such as water vapor, but copious amounts of cloud can masquerade as heavier atmospheres with water vapor in large abundance. Thus-far, observations of such planets, see for example \cite{Tsiaras_2019_k2-18}, \cite{Edwards_2020}, have made it near-impossible to distinguish between these light and heavy counterpart atmospheres, holding the true nature of these super-Earth - mini-Neptune atmospheres out of reach, \cite{changeat2021disentangling}. This is in large part due to the information content of observations that are possible today, in particular Hubble Space Telescope (HST) imagery, which lacks good enough spectral signal-to-noise, resolution and coverage.

As if to exacerbate an already difficult problem, currently available cross-section data in the exoplanet field is typically optimised for H$_2$-based atmospheres, with low mean-molecular-weight. Such cross-sections do not include contributions to molecular absorption such as, for example in the case of water, the water vapor continuum. The \textit{water continuum} is  the phenomenon of continuous water absorption which arises in the spectrum of water vapor, pervading the visible to microwave wavelengths \cite{mt_ckd,shine16}. The continuum is postulated to be due to two different mechanisms (which are not mutually exclusive); a contribution from the far-wings of water monomer lines, and a contribution from water dimers. The continuum is typically defined as the difference between the observed absorption by water vapor, and the contribution of the Lorentz-broadened water monomer lines with a 25 \cm\ wing cut-off. The continuum therefore contains a contribution from the monomer lines beyond 25 \cm, in addition to any non-Lorentzian far-wing contribution. The inclusion of this monomer contribution beyond 25 \cm\ means that the derived continuum is not independent of the spectral data used to define it; see e.g. \citep{shine16} for more details. The second possible explanation is that the continuum is in part caused by water dimers; complexes of two water molecules loosely bound by a shared hydrogen bond, which have a complex absorption spectrum which manifests as a continuum (see e.g. \citep{Ptashnik11b} for more details).

In this article we present simulated transmission spectra for two super-Earth planets: GJ 1214\,b and K2-18\,b. Using publicly available data from the CAVIAR experiment \citep{Ptashnik11a}, and cross-sections computed using line list data from the HITRAN database~\citep{jt691s,19TaKoRo} we include an absorption contribution from the water vapor self-continuum in our simulations with the retrieval suite TauREx 3.1 \citep{waldmann_2,waldmann_1,Al_Refaie_2021}.
We find that the differences in the simulated signal, as represented by the \textit{transit depth}, are non-negligible, and in comparison with simulated error bars for the near-future space-missions Ariel and JWST, these differences are detectable. The advent of high-resolution, low-signal-to-noise data-sets that these missions will provide make accurate modelling of heavy atmospheres vitally important. It is therefore imperative that alongside inclusion of effects like pressure-induced self-broadening \cite{19GhLi.broad,anisman21_a}, absorption contribution from the water continuum is included in the cross-section data used to analyse super-Earth and mini-Neptune atmospheres. 

\section{Methodology}
\subsection{Transmission spectroscopy}

If the orbital plane of a planet around its host star is aligned approximately parallel to our line of sight of the system (analagous to $90^{\circ}$ inclination), the planet will transit in front of its star. Assuming that there is no atmosphere, and that the planet is totally opaque to its incoming starlight, this transiting motion will cause a reduction in the amount of stellar flux we receive from the host star which is independent of wavelength. This change in detected light is known as the \emph{transit depth}, $\Delta F$, which is equal to the ratio between the surface area of the planet (as we view it, in 2D) and the surface area of the star. Since we assume both objects to be spherically symmetrical, this reduces to:
        \begin{equation} \Delta F = \frac{F_{out} - F_{in}}{F_{out}} = \left(\frac{R_{p}}{R_{*}}\right)^{2} 
        \label{eq:tran_depth_opaque}
        \end{equation}
        which gives a measure of the relative change in flux as the planet blocks its starlight. This provides us with an observable quantity with which we can quantify the size of the planet, if we know the stellar properties, which can be derived from models. 

        If the planet possesses an atmosphere, an envelope of gas which surrounds the planet which is maintained by the planet's gravitational force, the molecules present will absorb, scatter and reflect incoming starlight, in addition to thermally emitting photons. Owing to the varied and distinct spectral characteristics of different molecules, how opaque a certain atmosphere is to incoming stellar flux will vary significantly with wavelength. This information is described by the \emph{optical depth}, $\tau(\lambda)$, which is defined by Eq.~(\ref{attenuated_intensity}) below. Overall, regarding transmission spectroscopy, we can treat the atmosphere as a purely absorbing and single-scattering medium as a good (first-order) approximation for radiative transfer through the planetary atmosphere.

        Given an arbitrary path through the atmosphere for which radiation transmits with wavelength-dependent initial intensity $I_{\lambda}$, the transmitted radiance will be attenuated by absorption and scattering processes. We can denote this reduction in intensity as a function of path $ds$ as  $dI_{\lambda}/ds=-I_{\lambda}\sigma_{\lambda}\rho $, where $\sigma_{\lambda}$ is the total mass extinction cross section (the sum of the absorption and scattering cross sections) and $\rho$ is the density of the medium. Integrating up and using the fact that the optical depth as a function of atmospheric height is determined by summing the opacity contributions of all molecular species present, we recover the \emph{Beer-Bougert-Lambert Law}:
        \begin{equation}
            I_{\lambda}(z) = I_{\lambda}(0)e^{-\tau_{\lambda}(z)} \quad \text{with} \quad \tau_{\lambda}(z) = \sum_{m}\int_{z}^{z_{\infty}} \sigma_{m, \lambda}(z')\chi_{m}(z')\rho(z')dz',
            \label{attenuated_intensity}
        \end{equation} where $\chi_{m}$ and $\rho$ are the column density of a given molecular species and the number density of the atmosphere, respectively.

        We may now rewrite Eq.~(\ref{eq:tran_depth_opaque}) as:
        
        \begin{equation}
            \Delta F = \frac{F_{out} - F_{in}}{F_{out}} = \left(\frac{R_{p}+h_{\lambda}}{R_{*}}\right)^{2} \approx \frac{R_{p}^2+2R_{p}h_{\lambda}}{R_{*}^2} \qquad O(h_{\lambda})
            \label{eq:atmos_transit_depth}
        \end{equation}
        where we may describe the atmospheric height function as:
        
        \begin{equation}
            2R_{p}h_{z} = 2\int_{0}^{z_{max}} (R_{p}+z)(1-e^{-\tau_{\lambda}(z)}) dz,
            \label{eq:atmost_height_function}
        \end{equation} 
where $z_{max}$ denotes the height of the atmosphere.

Using this formalism, the transit depth of an atmosphere-bearing planet for any given wavelength may be calculated, using derived (temperature and wavelength dependent) \textit{cross-sections} for a given molecular species. If we populate a model atmosphere with a given temperature profile, pressure profile, chemical species abundances and a specified cloud distribution we may generate a transmission spectrum ($\Delta F$ vs. $\lambda$) for the atmosphere, thereby \textit{forward-modelling} it. To date, there is extensive literature pertaining to the collection of transmission data, alongside analysis of the generated transmission spectra, for a wide variety of exoplanets; from hot-Jupiters \cite{Skaf_2020} to habitable-zone super-Earths \cite{Edwards_2020}.  

\subsection{Simulating transmission spectra: TauREx 3.0}
In order to simulate forward models of transmission spectra for GJ 1214\,b and K2-18\,b, our analysis was performed using the retrieval suite TauREx 3.1. For the stellar parameters and the planet mass, we used the values from \cite{Harpsoe_GJ1214} and \cite{Benneke_2019}, as given in Table \ref{tab:params}. In our runs we assumed that the planets possess 10\% H$_2$O atmospheres with the remainder of the molecular abundance made up of H$_2$ and He, in the standard volume mixing ratio of He/H$_2$ = 0.17. We included in our simulations the absorption contribution of water vapor using cross-sections which we computed from the H$_2$O line list, from HITRAN 2016 using self-broadening coefficients. We used a fixed line wing cutoff of 25~cm$^{-1}$ in order to be compatible with the water continuum opacities used in this work. 
Additionally, we included the Collision Induced Absorption (CIA) from H$_2$-H$_2$ \citep{abel_h2-h2, fletcher_h2-h2} and H$_2$-He \citep{abel_h2-he}, as well as Rayleigh scattering for all molecules. Finally, our simulated atmospheres are cloud-free, isothermal and have molecular abundance profiles which are constant with altitude. For each of the planets two sets of spectra were generated: one using the standard HITRAN cross-sections, and another with the water continuum opacities included, as described. We note that not including absorption contribution from clouds in our forward run simulations, whilst not realistic, is nevertheless a deliberate choice. Alongside taking 10\% H$_2$O abundance, we maximise the modeled scale heights, and thus simply present a ``worst-case scenario'' in terms of the $\lambda$-dependent ppm differences in transit depth for both planets.
 
\subsection{CAVIAR lab data: water continuum}
\begin{figure}
    \centering
    \includegraphics[width=0.9\textwidth]{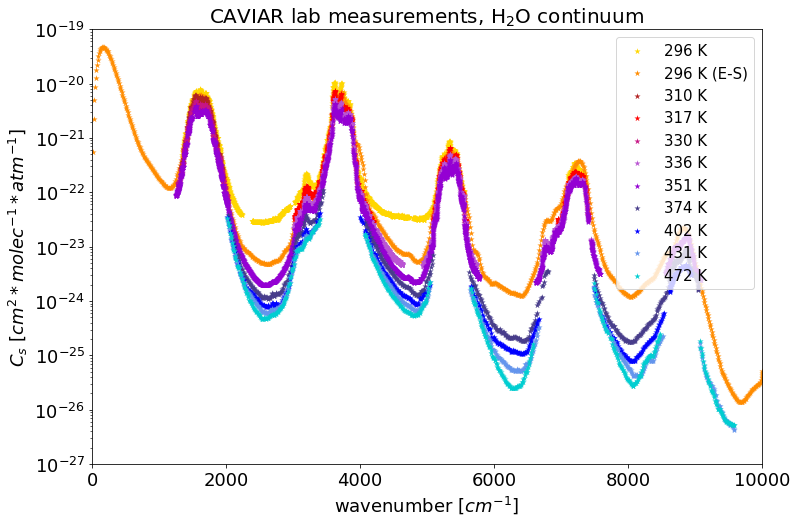}
    \includegraphics[width=0.9\textwidth]{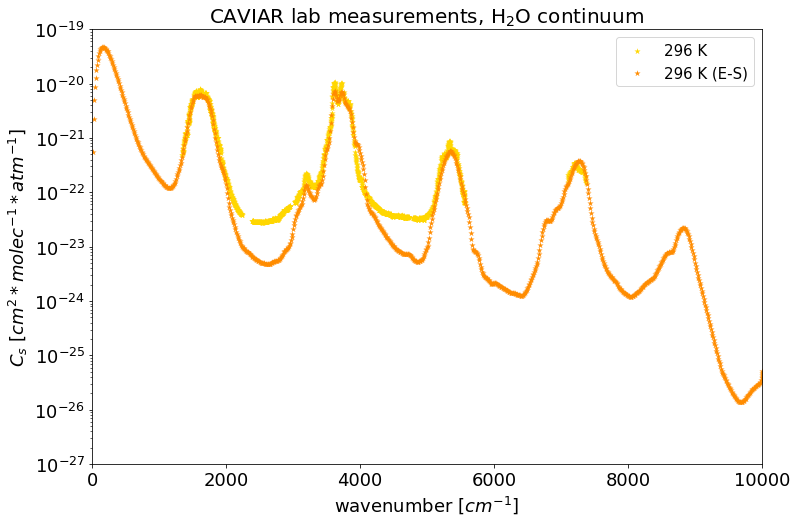}
    \caption{CAVIAR lab measurements, consolidated in Shine et al. 2016 \cite{shine16}, for H$_2$O self-continuum absorption, given as cross-section $C_{s}$ versus wave-number, as described in equation 1, for a range of temperatures, as shown. Top: raw data is displayed for all temperatures, bottom: only the 296 K datasets are shown, with the E-S continuum (described in text) over-plotted against the raw CAVIAR measurements.}
    \label{fig:caviar}
\end{figure}

In this work, we use data from the CAVIAR experiment \citep{Ptashnik11a} in the near-infrared windows between $2000 - 8000$ \cm, merged with MT\_CKD 3.2~\citep{mt_ckd} at other wavenumbers.  The continuum absorption coefficient of water vapor in air, $\alpha_{WC}$, is the sum of the self (WCS) and foreign (WCF) contributions,  i.e.:
\begin{align}
\alpha_{\rm WC}(\nu, T) &= \alpha_{\rm WCS}(\nu, T) + \alpha_{WCF}(\nu, T) \\
                    &= \frac{1}{k_{b}T}C_{S}(\nu, T){P^2_{{\rm H}_2 {\rm O}}}  + \frac{1}{k_{b}T}C_{F}(\nu, T){P_{{\rm H}_2 {\rm O}}}P_{F} \qquad (1)
\end{align}
\vspace{8mm}
where $k_b$ is the Boltzmann constant, $T$ is temperature, $P_{{\rm H}_2 {\rm O}}$ and $P_{F}$ are the water vapor and foreign gas (here air) partial pressures, respectively, and $C_{S}$ and $C_{F}$ represent the self- and foreign- continuum cross-sections, respectively, at a given temperature
$T$, following the definition in \cite{shine16}. In the standard ``HITRAN'' units adopted in this field, $C_{S}$ and $C_{F}$ are reported in $\text{\rm cm}^{2}\text{\rm molec}^{-1}\text{\rm atm}^{-1}$, $\alpha_{WC}$ is in $\text{\rm cm}^{-1}$, $P_{{\rm H}_2 {\rm O}}$ and $P_{F}$ are in atm, $T$ is in K and $k_{b} = 1.36 \times 10^{22} \quad \text{\rm molec}^{-1}\text{\rm cm}^3\text{\rm atm}\text{K}^{-1}$. Note that for comparison, the MT\_CKD values of $C_{S}$ and $C_{F}$, which are normalised to the number density at 1 atm and 296 K, should be multiplied by the factor $296/T$ (e.g. \cite{Ptashnik11a}). We do not consider the foreign continuum in this work, due to the composition of the atmospheres used, and therefore the results we present use the self-continuum only, and use ``water continuum'' to refer only to the self-continuum.

The CAVIAR/MT\_CKD data we use here between $0-10,000$ cm$^{-1},$ is displayed in Figure \ref{fig:caviar}. These data demonstrate a strong negative temperature dependence, broadly fitting the expected temperature dependence of the form $\exp(-D_0/T)$ (e.g. \cite{Ptashnik11b}), where $D_0$ relates to the dissociation energy of the water dimer. Recent work (e.g. \cite{PTASHNIK201997}) has motivated a re-examination of the room temperature data from CAVIAR in the 2.1 and 4 \um\ windows. This is in part due to the the possibility of adsorption on the kind of gold mirror coatings used in e.g. \cite{Ptashnik11a,ptashnik2015b}. This effect is particularly pronounced at room temperature, where the relative humidity is higher than at high temperature. Elsey {\it et al.} \cite{elsey20} demonstrate that the room temperature CAVIAR data deviate from the temperature dependence implied by the higher-temperature measurements (which have lower uncertainty) by up to a factor of 5 in these windows at 296 K.

While it cannot be ruled out that this deviation is a physical effect (e.g. due to a transition between quasi-bound and true-bound dimers), this effect is not observed in measurements using a cavity ring-down technique \cite{vent2015}, which have smaller uncertainties; Elsey {\it et al.} \cite{elsey20} describe this issue in more detail. Conversely, extrapolating high-$T$ CAVIAR data to room temperature using this $\exp(-D_0/T)$ dependence leads to very good agreement with the cavity ring-down observations in the 2.1 and 4 \um\ windows, although large unresolved discrepancies between laboratory measurements remain at 1.6 \um. We therefore use these extrapolated CAVIAR data in this work (henceforth the Elsey-Shine or E-S continuuum). The E-S continuum also includes in-band observations from \cite{PTASHNIK201997} and \cite{Paynter09}. To this effect, in the second panel of Figure \ref{fig:caviar} we over-plot the raw CAVIAR experimental data (yellow) with the thermally-extrapolated data (orange). For our analysis of GJ 1214\,b, we use the 472 K observations from CAVIAR, while for K2-18\,b we extrapolate the available data to 296 K excluding the room temperature observations (i.e. the orange line in Fig. 1, panel 2).

To simulate the atmospheres of our chosen planets, including absorption contribution from the continuum, we simply build an additional class inside TauREx 3.1. Using the data-sets described in Section 2.2, we then add this extra absorption to the list of contributions (see Section 2.1), using the cross-sections for water vapor computed using HITRAN, with a fixed line wing cutoff of 25~cm$^{-1}$, as described.

\subsection{Ariel \& JWST error-bars}
During its primary mission, Ariel will survey the atmospheres of 1000 exoplanets \citep{edwards_ariel} while JWST could observe up to 150 over the 5-year mission lifetime \citep{cowan}. We generate error bars for the simulated spectra using ArielRad \citep{mugnai} and ExoWebb, which is constructed using the methods described in \cite{Edwards_2021}, in order to compare these with the absolute differences between including the water continuum and not in the planetary transmission spectra, to investigate the detectability of continuum-induced differences in transit depth. For JWST we modeled observations with NIRISS GR700XD (0.8 - 2.8 \um) and NIRSpec G395M (2.9 - 5.3 $\mu m$), whilst for Ariel, which provides simultaneous coverage from 0.5 - 7.8 \um, we simulated error bars at Tier 2 resolution. 

\section{Results}

\begin{table}
\centering
\begin{tabular}{cccc}
\hline\hline
\multicolumn{3}{c}{stellar \& planetary parameters}\\ \hline
    parameter & K2-18\,b & GJ 1214\,b \\\hline
    $T_{*}$ [K]  & 3457 & 3026 &\\
    $R_{*}$ [$R_\odot$]  & 0.4445 & 0.220&\\
    $M_{*}$ [$M_\odot$]  & 0.4951 & 0.176 &\\
    $M_{p}$ [$M_{\bigoplus}$] & 8.63 & 6.26 &\\
    $R_{p}$ [$R_{\bigoplus}$] & 2.61 & 2.85&\\
    $P_{orbital}$ [days] & 32.94 & 1.58\\
    \end{tabular}
    \begin{tabular}{cccc}
  \hline \hline
\multicolumn{4}{c}{forward model parameters}\\ \hline
\mbox{parameter}       & \mbox{K2-18\,b}   & \mbox{GJ 1214\,b} & \mbox{type} \\
\hline
\mbox{$P_{clouds}$}   & \mbox{None}  & \mbox{None} & \mbox{opaque} \\
\mbox{$T$ [K]}       & \mbox{$296$} & \mbox{$472$}  & \mbox{iso.}\\
\mbox{$V_{{\text{H}_2\text{O}}}$}  & 0.1  & 0.1 & \mbox{fill}\\
\hline \hline
 \end{tabular}
 \caption{Top: stellar and planetary parameters for K2-18\,b \& GJ 1214\,b, for input into TauREx 3.1, derived from \cite{Benneke_2019}, \cite{Gillon_2014} and \cite{Harpsoe_GJ1214}; Bottom: list of the forward-modeled parameters, their values and the scaling used.}
\label{tab:params}
 \end{table}

\subsection{GJ 1214 b: CAVIAR continuum}

\begin{figure*}
    \centering
    \includegraphics[width=0.635\textwidth]{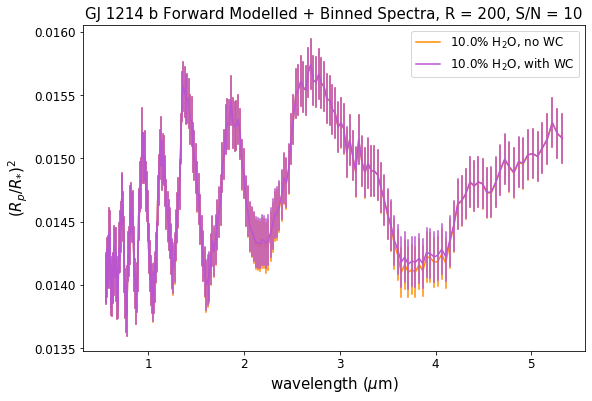}
     \includegraphics[width=0.6\textwidth]{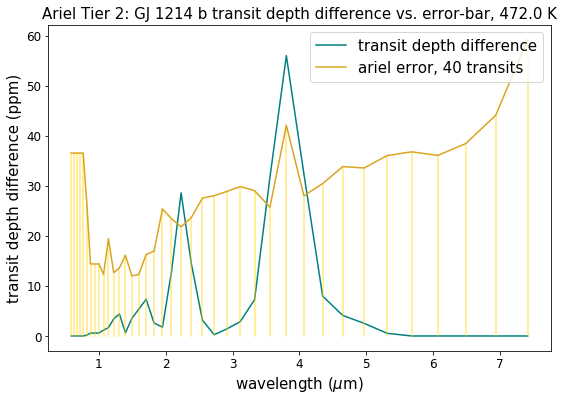}
    \includegraphics[width=0.6\textwidth]{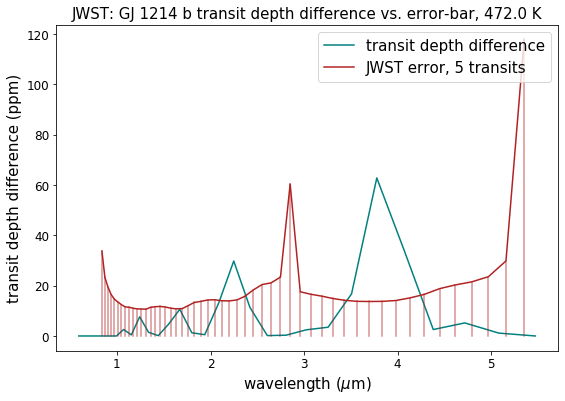}
     
    \caption{Top: over-plotted transmission spectra comparing inclusion of the water continuum with omitting it. The purple and orange data-sets are subtracted to yield the absolute differences in transit depth, given in blue in the second and third panels. Middle: transit depth difference vs. Ariel Tier 2 error-bar and bottom: vs. JWST error-bar for specified no. of transits. All continuum data for 472 K is raw CAVIAR data.}
    \label{fig:gj1214}
\end{figure*}

Using the parameters listed in Table \ref{tab:params}, we simulate two transmission spectra for GJ 1214\,b, one with absorption from the water continuum included using the 472 K CAVIAR data, and one without. This is the warmest available data-set; hence we simulate a GJ 1214\,b-like analogue planet, since the equilibrium temperature of GJ 1214\,b inferred from observations \cite{Gillon_2014} is 604 K. We generate both spectra at a native resolution of 15,000, before binning them down to a nominal resolution and signal-to-noise of 200 and 10, respectively (enabling eventual comparison with Ariel and JWST error bars). We present the results in the top panel of Figure \ref{fig:gj1214}. Upon subtracting the spectra in the top panel, for Ariel Tier 2 binning down to a resolution of 15, we obtain the middle panel, displaying the absolute difference in transit depth versus wavelength, in ppm, compared with the simulated error-bar for 40 transits. The bottom panel presents the analogous plot for 5 transits with the JWST NIRSpec and NIRISS instruments. Hence, we show that with a finite, reasonable number of observations, a ppm difference of up to 60 ppm induced by including the water continuum should be observable; requiring 5 and 40 transits with JWST and Ariel, respectively. In particular, the transit depth difference competes with the magnitude of the Ariel error-bar, whilst at $4\mu m$ a $\sim3\sigma$ significance is found in terms of detectability with JWST.

\subsection{K2-18 b: E-S continuum}
\begin{figure}
    \centering
    \includegraphics[width=0.635\textwidth]{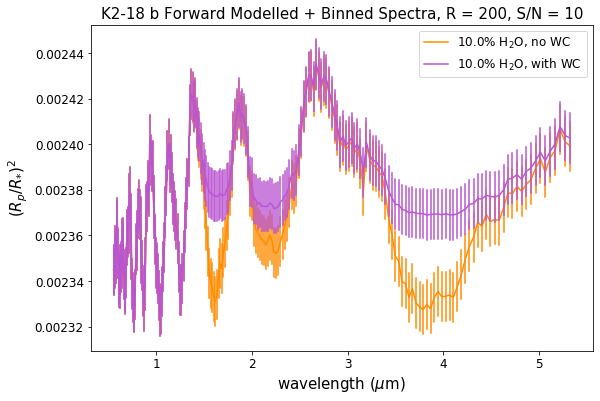}
    \includegraphics[width=0.6\textwidth]{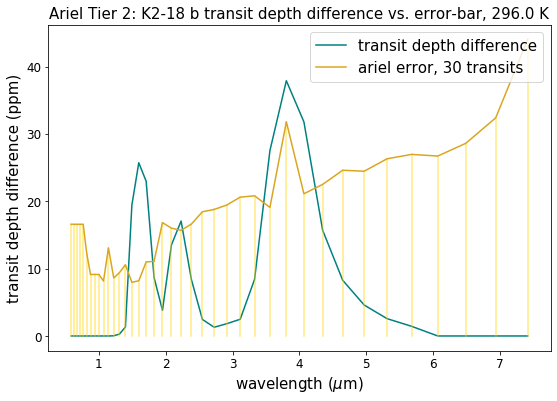}
    \includegraphics[width=0.6\textwidth]{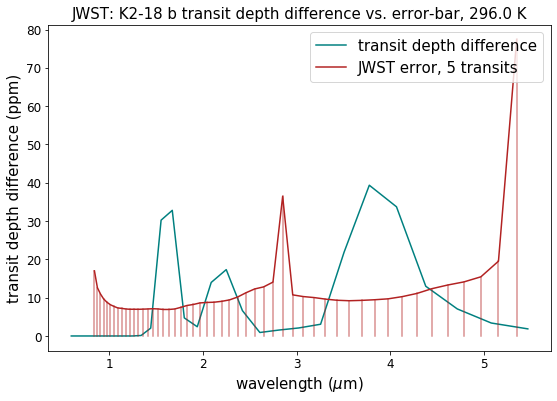}

    \caption{Top: over-plotted transmission spectra comparing inclusion of the water continuum with omitting it. The purple and orange data-sets are subtracted to yield the absolute differences in transit depth, given in blue in the second and third panels. Middle: transit depth difference vs. Ariel Tier 2 error-bar and bottom: vs. JWST error-bar for specified no. of transits. All continuum data for 296 K is using the E-S continuum.}
    \label{fig:k218}
\end{figure}

Using the parameters listed in Table \ref{tab:params}, we simulate two transmission spectra for K2-18\,b, one with absorption contribution from the water continuum included using the E-S continuum for 296 K, and one without. We generate both spectra at a native resolution of 15,000, before binning them down to a nominal resolution and signal-to-noise of 200 and 10, respectively (enabling eventual comparison with Ariel and JWST error-bars). We present the results in the top panel of Figure \ref{fig:k218}. Upon subtracting the spectra in the top panel, for Ariel Tier 2 binning down to a resolution of 15, we obtain the middle panel, displaying the absolute differences in transit depth versus wavelength, in ppm, compared with the simulated error-bar for 30 transits. The bottom panel presents the analogous plot for 5 transits with the JWST NIRSpec and NIRISS instruments. Hence, we show that with finite observations, a ppm difference of up to 40 ppm induced by including the water continuum should be reasonably observable; requiring 5 and 30 transits with JWST and Ariel, respectively. In particular, due to the cooler temperature, a significant water continuum spectral feature is found at $\sim 1.8 \mu m$. Hence, in addition to the water-continuum-induced differences in transit depth being impactful for the transmission spectra of K2-18 b, these differences also have a $\sim 3\sigma$ significance in terms of detectability for both Ariel \& JWST.

\section{Discussion}

We have demonstrated that by including absorption contribution due to the water vapor continuum in our simulations of small planet transmission spectra significantly impacts the transit depth. A caveat to our analysis is that there is significant uncertainty in the true strength of the self-continuum at room temperature. Observations at room temperature have inherently larger uncertainty than observations at elevated temperatures; since the saturation vapor pressure is significantly lower, observations must be made at lower vapor pressures. While the E-S self-continuum used in this work has good agreement with some other available data-sets in the 2.1 and 4 \um\ windows, in the centre of the 1.6 \um\ window differences between different data-sets when extrapolated to room temperature are up to an order of magnitude or greater \cite{shine16,elsey20}.

The decision to use the E-S continuum rather than the raw CAVIAR data at 296 K is motivated by the agreement of E-S with the available cavity ring-down (CRDS) data, which have low-stated uncertainties, the large stated uncertainties in the room-$T$ CAVIAR data, and the possibility of adsorption on the gold mirror coatings used in CAVIAR (which is not accounted for in the uncertainty budget). However, it is also possible that a similar issue could have affected the CRDS measurements (e.g. \cite{Serdyukov:16}), and that the observed deviation from the $\exp(-D_0/T)$ temperature dependence in CAVIAR is physical. More work (both experimental and theoretical) is therefore required to understand the water vapor self-continuum and its temperature dependence.

\section{Conclusion}

In conclusion, we have illustrated the necessity for the absorption contribution induced by the water vapor continuum to be included in the analysis of super-Earth/mini-Neptune transmission spectra. We show that the simulated transit depth can be altered by as much as 60 ppm, with the ppm differences for small, temperate exoplanets GJ 1214\,b and K2-18\,b sitting above the noise level for a reasonable number of observations with the near-future space telescopes JWST and Ariel. The CAVIAR data utilised in this work exhibits an inverse temperature dependence, meaning that colder atmospheres are more significantly impacted by absorption from the water continuum, hence omitting its contribution is far more impactful for low-temperature atmospheres. Thus, it is essential that the exoplanet spectroscopy field adopts the use of more adaptive cross-sections, built as functions not only of temperature and pressure but also of molecular abundance; in particular including the effect of the water continuum specifically for the case of H$_2$O.

\section{Acknowledgements}
\textit{Funding:} We acknowledge funding through the ERC Consolidator grant ExoAI (GA 758892), the UK Space Agency grant UKSA (ST/W00254X/1) and Advance grant ExoMolHD (883830), as well as STFC grants:
\\
ST/P000282/1, ST/P002153/1, ST/S002634/1 and ST/T001836/1, 
\\
and NERC grants NE/R009848/1
and NE/T000767/1.
\\

This work utilised resources provided by the Cambridge Service for Data Driven Discovery (CSD3) operated by the University of Cambridge Research Computing Service (www.csd3.cam.ac.uk), provided by Dell EMC and Intel using Tier-2 funding from the Engineering and Physical Sciences Research Council (capital grant EP/P020259/1), and DiRAC funding from the Science and Technology Facilities Council (www.dirac.ac.uk).

\bibliographystyle{elsarticle-num-names} 
\bibliography{elsarticle-template-num}

\begin{thebibliography}{34}
\expandafter\ifx\csname natexlab\endcsname\relax\def\natexlab#1{#1}\fi
\providecommand{\url}[1]{\texttt{#1}}
\providecommand{\href}[2]{#2}
\providecommand{\path}[1]{#1}
\providecommand{\DOIprefix}{doi:}
\providecommand{\ArXivprefix}{arXiv:}
\providecommand{\URLprefix}{URL: }
\providecommand{\Pubmedprefix}{pmid:}
\providecommand{\doi}[1]{\href{http://dx.doi.org/#1}{\path{#1}}}
\providecommand{\Pubmed}[1]{\href{pmid:#1}{\path{#1}}}
\providecommand{\bibinfo}[2]{#2}
\ifx\xfnm\relax \def\xfnm[#1]{\unskip,\space#1}\fi
\bibitem[{Kreidberg et~al.(2014)Kreidberg, Bean, D\'{e}sert, Benneke, Deming,
  Stevenson, Seager, Berta-Thompson, Seifahrt, and Homeier}]{Kreidberg_2014}
\bibinfo{author}{L.~Kreidberg}, \bibinfo{author}{J.~L. Bean},
  \bibinfo{author}{J.-M. D\'{e}sert}, \bibinfo{author}{B.~Benneke},
  \bibinfo{author}{D.~Deming}, \bibinfo{author}{K.~B. Stevenson},
  \bibinfo{author}{S.~Seager}, \bibinfo{author}{Z.~Berta-Thompson},
  \bibinfo{author}{A.~Seifahrt}, \bibinfo{author}{D.~Homeier},
\newblock \bibinfo{title}{{Clouds in the atmosphere of the super-Earth
  exoplanet GJ1214b}},
\newblock \bibinfo{journal}{Nature} \bibinfo{volume}{505}
  (\bibinfo{year}{2014}) \bibinfo{pages}{69–72}.
  \DOIprefix\doi{10.1038/nature12888}.
\bibitem[{Libby-Roberts et~al.(2021)Libby-Roberts, Berta-Thompson,
  Diamond-Lowe, Gully-Santiago, Irwin, Kempton, Rackham, Charbonneau, Desert,
  Dittmann, Hofmann, Morley, and Newton}]{LR_2020}
\bibinfo{author}{J.~Libby-Roberts}, \bibinfo{author}{Z.~Berta-Thompson},
  \bibinfo{author}{H.~Diamond-Lowe}, \bibinfo{author}{M.~Gully-Santiago},
  \bibinfo{author}{J.~Irwin}, \bibinfo{author}{E.~Kempton},
  \bibinfo{author}{B.~Rackham}, \bibinfo{author}{D.~Charbonneau},
  \bibinfo{author}{J.-M. Desert}, \bibinfo{author}{J.~Dittmann},
  \bibinfo{author}{R.~Hofmann}, \bibinfo{author}{C.~Morley},
  \bibinfo{author}{E.~Newton},
\newblock \bibinfo{title}{The featureless hst/wfc3 transmission spectrum of the
  rocky exoplanet gj 1132b: No evidence for a cloud-free primordial atmosphere
  and constraints on starspot contamination},
\newblock \bibinfo{journal}{Astrophysical Journal}  (\bibinfo{year}{2021}).
\bibitem[{Swain et~al.(2021)Swain, Estrela, Roudier, Sotin, Rimmer, Valio,
  West, Pearson, Huber-Feely, and Zellem}]{Swain_2021}
\bibinfo{author}{M.~R. Swain}, \bibinfo{author}{R.~Estrela},
  \bibinfo{author}{G.~M. Roudier}, \bibinfo{author}{C.~Sotin},
  \bibinfo{author}{P.~B. Rimmer}, \bibinfo{author}{A.~Valio},
  \bibinfo{author}{R.~West}, \bibinfo{author}{K.~Pearson},
  \bibinfo{author}{N.~Huber-Feely}, \bibinfo{author}{R.~T. Zellem},
\newblock \bibinfo{title}{Detection of an atmosphere on a rocky exoplanet},
\newblock \bibinfo{journal}{The Astronomical Journal} \bibinfo{volume}{161}
  (\bibinfo{year}{2021}) \bibinfo{pages}{213}.
  \DOIprefix\doi{10.3847/1538-3881/abe879}.
\bibitem[{Tsiaras et~al.(2019)Tsiaras, Waldmann, Tinetti, Tennyson, and
  Yurchenko}]{Tsiaras_2019_k2-18}
\bibinfo{author}{A.~Tsiaras}, \bibinfo{author}{I.~P. Waldmann},
  \bibinfo{author}{G.~Tinetti}, \bibinfo{author}{J.~Tennyson},
  \bibinfo{author}{S.~N. Yurchenko},
\newblock \bibinfo{title}{{Water vapour in the atmosphere of the habitable-zone
  eight-Earth-mass planet K2-18 b}},
\newblock \bibinfo{journal}{Nature Astronomy}  (\bibinfo{year}{2019}).
  \URLprefix \url{http://dx.doi.org/10.1038/s41550-019-0878-9}.
  \DOIprefix\doi{10.1038/s41550-019-0878-9}.
\bibitem[{Edwards et~al.(2020)Edwards, Changeat, Mori, Anisman, Morvan, Yip,
  Tsiaras, Al-Refaie, Waldmann, and Tinetti}]{Edwards_2020}
\bibinfo{author}{B.~Edwards}, \bibinfo{author}{Q.~Changeat},
  \bibinfo{author}{M.~Mori}, \bibinfo{author}{L.~O. Anisman},
  \bibinfo{author}{M.~Morvan}, \bibinfo{author}{K.~H. Yip},
  \bibinfo{author}{A.~Tsiaras}, \bibinfo{author}{A.~Al-Refaie},
  \bibinfo{author}{I.~Waldmann}, \bibinfo{author}{G.~Tinetti},
\newblock \bibinfo{title}{{Hubble WFC3 Spectroscopy of the Habitable-zone
  Super-Earth LHS 1140 b}},
\newblock \bibinfo{journal}{The Astronomical Journal} \bibinfo{volume}{161}
  (\bibinfo{year}{2020}) \bibinfo{pages}{44}. \URLprefix
  \url{http://dx.doi.org/10.3847/1538-3881/abc6a5}.
  \DOIprefix\doi{10.3847/1538-3881/abc6a5}.
\bibitem[{Changeat et~al.(2021)Changeat, Edwards, Al-Refaie, Tsiaras, Waldmann,
  and Tinetti}]{changeat2021disentangling}
\bibinfo{author}{Q.~Changeat}, \bibinfo{author}{B.~Edwards},
  \bibinfo{author}{A.~F. Al-Refaie}, \bibinfo{author}{A.~Tsiaras},
  \bibinfo{author}{I.~P. Waldmann}, \bibinfo{author}{G.~Tinetti},
\newblock \bibinfo{title}{Disentangling atmospheric compositions of k2-18 b
  with next generation facilities},
\newblock \bibinfo{journal}{Experimental Astronomy}  (\bibinfo{year}{2021}).
\bibitem[{Mlawer et~al.(2012)Mlawer, Payne, Moncet, Delamere, Alvarado, and
  Tobin}]{mt_ckd}
\bibinfo{author}{E.~J. Mlawer}, \bibinfo{author}{V.~H. Payne},
  \bibinfo{author}{J.~L. Moncet}, \bibinfo{author}{J.~S. Delamere},
  \bibinfo{author}{M.~J. Alvarado}, \bibinfo{author}{D.~C. Tobin},
\newblock \bibinfo{title}{Development and recent evaluation of the {MT\_CKD}
  model of continuum absorption},
\newblock \bibinfo{journal}{Phil. Trans. Roy. Soc. A} \bibinfo{volume}{370}
  (\bibinfo{year}{2012}) \bibinfo{pages}{2520--2556}.
  \DOIprefix\doi{10.1098/rsta.2011.0295}.
\bibitem[{Shine et~al.(2016)Shine, Campargue, Mondelain, McPheat, Ptashnik, and
  Weidmann}]{shine16}
\bibinfo{author}{K.~P. Shine}, \bibinfo{author}{A.~Campargue},
  \bibinfo{author}{D.~Mondelain}, \bibinfo{author}{R.~A. McPheat},
  \bibinfo{author}{I.~V. Ptashnik}, \bibinfo{author}{D.~Weidmann},
\newblock \bibinfo{title}{The water vapour continuum in near-infrared windows
  – current understanding and prospects for its inclusion in spectroscopic
  databases},
\newblock \bibinfo{journal}{Journal of Molecular Spectroscopy}
  \bibinfo{volume}{327} (\bibinfo{year}{2016}) \bibinfo{pages}{193--208}.
  \DOIprefix\doi{https://doi.org/10.1016/j.jms.2016.04.011}.
\bibitem[{Ptashnik et~al.(2011{\natexlab{a}})Ptashnik, Shine, and
  Vigasin}]{Ptashnik11b}
\bibinfo{author}{I.~V. Ptashnik}, \bibinfo{author}{K.~P. Shine},
  \bibinfo{author}{A.~A. Vigasin},
\newblock \bibinfo{title}{{Water vapour self-continuum and water dimers. 1.
  Analysis of recent work}},
\newblock \bibinfo{journal}{Journal of Quantitative Spectroscopy and Radiative
  Transfer} \bibinfo{volume}{112} (\bibinfo{year}{2011}{\natexlab{a}})
  \bibinfo{pages}{1286--1303}. \DOIprefix\doi{10.1016/j.jqsrt.2011.01.012}.
\bibitem[{Ptashnik et~al.(2011{\natexlab{b}})Ptashnik, McPheat, Shine, Smith,
  and Williams}]{Ptashnik11a}
\bibinfo{author}{I.~V. Ptashnik}, \bibinfo{author}{R.~A. McPheat},
  \bibinfo{author}{K.~P. Shine}, \bibinfo{author}{K.~M. Smith},
  \bibinfo{author}{R.~G. Williams},
\newblock \bibinfo{title}{{Water vapor self-continuum absorption in
  near-infrared windows derived from laboratory measurements}},
\newblock \bibinfo{journal}{Journal of Geophysical Research}
  \bibinfo{volume}{116} (\bibinfo{year}{2011}{\natexlab{b}})
  \bibinfo{pages}{D16305}.
\bibitem[{Gordon and {et al.}(2017)}]{jt691s}
\bibinfo{author}{I.~E. Gordon}, \bibinfo{author}{{et al.}},
\newblock \bibinfo{title}{{The \textit{ HITRAN} 2016 molecular spectroscopic
  database}},
\newblock \bibinfo{journal}{J. Quant. Spectrosc. Radiat. Transf.}
  \bibinfo{volume}{203} (\bibinfo{year}{2017}) \bibinfo{pages}{3--69}.
  \DOIprefix\doi{10.1016/j.jqsrt.2017.06.038}.
\bibitem[{Tan et~al.(2019)Tan, Kochanov, Rothman, and Gordon}]{19TaKoRo}
\bibinfo{author}{Y.~Tan}, \bibinfo{author}{R.~V. Kochanov},
  \bibinfo{author}{L.~S. Rothman}, \bibinfo{author}{I.~E. Gordon},
\newblock \bibinfo{title}{Introduction of water-vapor broadening parameters and
  their temperature-dependent exponents into the {HITRAN} database: Part
  {I—CO2, N2O, CO, CH4, O2, NH3, and H2S}},
\newblock \bibinfo{journal}{Journal of Geophysical Research: Atmospheres}
  \bibinfo{volume}{124} (\bibinfo{year}{2019}) \bibinfo{pages}{11580--11594}.
  \DOIprefix\doi{10.1029/2019JD030929}.
\bibitem[{{Waldmann} et~al.(2015{\natexlab{a}}){Waldmann}, {Rocchetto},
  {Tinetti}, {Barton}, {Yurchenko}, and {Tennyson}}]{waldmann_2}
\bibinfo{author}{I.~P. {Waldmann}}, \bibinfo{author}{M.~{Rocchetto}},
  \bibinfo{author}{G.~{Tinetti}}, \bibinfo{author}{E.~J. {Barton}},
  \bibinfo{author}{S.~N. {Yurchenko}}, \bibinfo{author}{J.~{Tennyson}},
\newblock \bibinfo{title}{{Tau-REx II: Retrieval of Emission Spectra}},
\newblock \bibinfo{journal}{The Astrophysical Journal} \bibinfo{volume}{813}
  (\bibinfo{year}{2015}{\natexlab{a}}) \bibinfo{pages}{13}.
  \DOIprefix\doi{10.1088/0004-637X/813/1/13}.
\bibitem[{{Waldmann} et~al.(2015{\natexlab{b}}){Waldmann}, {Tinetti},
  {Rocchetto}, {Barton}, {Yurchenko}, and {Tennyson}}]{waldmann_1}
\bibinfo{author}{I.~P. {Waldmann}}, \bibinfo{author}{G.~{Tinetti}},
  \bibinfo{author}{M.~{Rocchetto}}, \bibinfo{author}{E.~J. {Barton}},
  \bibinfo{author}{S.~N. {Yurchenko}}, \bibinfo{author}{J.~{Tennyson}},
\newblock \bibinfo{title}{{Tau-REx I: A Next Generation Retrieval Code for
  Exoplanetary Atmospheres}},
\newblock \bibinfo{journal}{The Astrophysical Journal} \bibinfo{volume}{802}
  (\bibinfo{year}{2015}{\natexlab{b}}) \bibinfo{pages}{107}.
  \DOIprefix\doi{10.1088/0004-637X/802/2/107}.
\bibitem[{Al-Refaie et~al.(2021)Al-Refaie, Changeat, Waldmann, and
  Tinetti}]{Al_Refaie_2021}
\bibinfo{author}{A.~F. Al-Refaie}, \bibinfo{author}{Q.~Changeat},
  \bibinfo{author}{I.~P. Waldmann}, \bibinfo{author}{G.~Tinetti},
\newblock \bibinfo{title}{{TauREx} 3: A fast, dynamic, and extendable framework
  for retrievals},
\newblock \bibinfo{journal}{The Astrophysical Journal} \bibinfo{volume}{917}
  (\bibinfo{year}{2021}) \bibinfo{pages}{37}. \URLprefix
  \url{https://doi.org/10.3847/1538-4357/ac0252}.
  \DOIprefix\doi{10.3847/1538-4357/ac0252}.
\bibitem[{Gharib-Nezhad and Line(2019)}]{19GhLi.broad}
\bibinfo{author}{E.~Gharib-Nezhad}, \bibinfo{author}{M.~R. Line},
\newblock \bibinfo{title}{The influence of {H$_2$O} pressure broadening in
  high-metallicity exoplanet atmospheres},
\newblock \bibinfo{journal}{Astrophys. J.} \bibinfo{volume}{872}
  (\bibinfo{year}{2019}) \bibinfo{pages}{27}.
  \DOIprefix\doi{10.3847/1538-4357/aafb7b}.
\bibitem[{Anisman et~al.(prep)Anisman, Chubb, Changeat, and
  Tinetti}]{anisman21_a}
\bibinfo{author}{L.~O. Anisman}, \bibinfo{author}{K.~L. Chubb},
  \bibinfo{author}{Q.~Changeat}, \bibinfo{author}{G.~Tinetti},
\newblock \bibinfo{title}{{Cross-sections for heavy atmospheres: H$_2$O
  self-broadening}},
\newblock \bibinfo{journal}{JQSRT}  (\bibinfo{year}{in prep}).
\bibitem[{Skaf et~al.(2020)Skaf, Bieger, Edwards, Changeat, Morvan, Kiefer,
  Blain, Zingales, Poveda, Al-Refaie, and et~al.}]{Skaf_2020}
\bibinfo{author}{N.~Skaf}, \bibinfo{author}{M.~F. Bieger},
  \bibinfo{author}{B.~Edwards}, \bibinfo{author}{Q.~Changeat},
  \bibinfo{author}{M.~Morvan}, \bibinfo{author}{F.~Kiefer},
  \bibinfo{author}{D.~Blain}, \bibinfo{author}{T.~Zingales},
  \bibinfo{author}{M.~Poveda}, \bibinfo{author}{A.~Al-Refaie},
  \bibinfo{author}{et~al.},
\newblock \bibinfo{title}{Ares. ii. characterizing the hot jupiters wasp-127 b,
  wasp-79 b, and wasp-62b with the hubble space telescope},
\newblock \bibinfo{journal}{The Astronomical Journal} \bibinfo{volume}{160}
  (\bibinfo{year}{2020}) \bibinfo{pages}{109}. \URLprefix
  \url{http://dx.doi.org/10.3847/1538-3881/ab94a3}.
  \DOIprefix\doi{10.3847/1538-3881/ab94a3}.
\bibitem[{{Harps{\o}e} et~al.(2013){Harps{\o}e}, {Hardis}, {Hinse},
  {J{\o}rgensen}, {Mancini}, {Southworth}, {Alsubai}, {Bozza}, {Browne},
  {Burgdorf}, {Calchi Novati}, {Dodds}, {Dominik}, {Fang}, {Finet}, {Gerner},
  {Gu}, {Hundertmark}, {Jessen-Hansen}, {Kains}, {Kerins}, {Kjeldsen},
  {Liebig}, {Lund}, {Lundkvist}, {Mathiasen}, {Nesvorn{\'y}}, {Nikolov},
  {Penny}, {Proft}, {Rahvar}, {Ricci}, {Sahu}, {Scarpetta}, {Sch{\"a}fer},
  {Sch{\"o}nebeck}, {Snodgrass}, {Skottfelt}, {Surdej}, {Tregloan-Reed}, and
  {Wertz}}]{Harpsoe_GJ1214}
\bibinfo{author}{K.~B.~W. {Harps{\o}e}}, \bibinfo{author}{S.~{Hardis}},
  \bibinfo{author}{T.~C. {Hinse}}, \bibinfo{author}{U.~G. {J{\o}rgensen}},
  \bibinfo{author}{L.~{Mancini}}, \bibinfo{author}{J.~{Southworth}},
  \bibinfo{author}{K.~A. {Alsubai}}, \bibinfo{author}{V.~{Bozza}},
  \bibinfo{author}{P.~{Browne}}, \bibinfo{author}{M.~J. {Burgdorf}},
  \bibinfo{author}{S.~{Calchi Novati}}, \bibinfo{author}{P.~{Dodds}},
  \bibinfo{author}{M.~{Dominik}}, \bibinfo{author}{X.~S. {Fang}},
  \bibinfo{author}{F.~{Finet}}, \bibinfo{author}{T.~{Gerner}},
  \bibinfo{author}{S.~H. {Gu}}, \bibinfo{author}{M.~{Hundertmark}},
  \bibinfo{author}{J.~{Jessen-Hansen}}, \bibinfo{author}{N.~{Kains}},
  \bibinfo{author}{E.~{Kerins}}, \bibinfo{author}{H.~{Kjeldsen}},
  \bibinfo{author}{C.~{Liebig}}, \bibinfo{author}{M.~N. {Lund}},
  \bibinfo{author}{M.~{Lundkvist}}, \bibinfo{author}{M.~{Mathiasen}},
  \bibinfo{author}{D.~{Nesvorn{\'y}}}, \bibinfo{author}{N.~{Nikolov}},
  \bibinfo{author}{M.~T. {Penny}}, \bibinfo{author}{S.~{Proft}},
  \bibinfo{author}{S.~{Rahvar}}, \bibinfo{author}{D.~{Ricci}},
  \bibinfo{author}{K.~C. {Sahu}}, \bibinfo{author}{G.~{Scarpetta}},
  \bibinfo{author}{S.~{Sch{\"a}fer}}, \bibinfo{author}{F.~{Sch{\"o}nebeck}},
  \bibinfo{author}{C.~{Snodgrass}}, \bibinfo{author}{J.~{Skottfelt}},
  \bibinfo{author}{J.~{Surdej}}, \bibinfo{author}{J.~{Tregloan-Reed}},
  \bibinfo{author}{O.~{Wertz}},
\newblock \bibinfo{title}{{The transiting system GJ1214: high-precision
  defocused transit observations and a search for evidence of transit timing
  variation}},
\newblock \bibinfo{journal}{Astronomy \& Astrophysics} \bibinfo{volume}{549}
  (\bibinfo{year}{2013}) \bibinfo{pages}{A10}.
  \DOIprefix\doi{10.1051/0004-6361/201219996}.
\bibitem[{Benneke et~al.(2019)Benneke, Wong, Piaulet, Knutson, Lothringer,
  Morley, Crossfield, Gao, Greene, Dressing, and et~al.}]{Benneke_2019}
\bibinfo{author}{B.~Benneke}, \bibinfo{author}{I.~Wong},
  \bibinfo{author}{C.~Piaulet}, \bibinfo{author}{H.~A. Knutson},
  \bibinfo{author}{J.~Lothringer}, \bibinfo{author}{C.~V. Morley},
  \bibinfo{author}{I.~J.~M. Crossfield}, \bibinfo{author}{P.~Gao},
  \bibinfo{author}{T.~P. Greene}, \bibinfo{author}{C.~Dressing},
  \bibinfo{author}{et~al.},
\newblock \bibinfo{title}{Water vapor and clouds on the habitable-zone
  sub-neptune exoplanet k2-18b},
\newblock \bibinfo{journal}{The Astrophysical Journal} \bibinfo{volume}{887}
  (\bibinfo{year}{2019}) \bibinfo{pages}{L14}. \URLprefix
  \url{http://dx.doi.org/10.3847/2041-8213/ab59dc}.
  \DOIprefix\doi{10.3847/2041-8213/ab59dc}.
\bibitem[{Abel et~al.(2011)Abel, Frommhold, Li, and Hunt}]{abel_h2-h2}
\bibinfo{author}{M.~Abel}, \bibinfo{author}{L.~Frommhold},
  \bibinfo{author}{X.~Li}, \bibinfo{author}{K.~L. Hunt},
\newblock \bibinfo{title}{Collision-induced absorption by {H$_2$} pairs: From
  hundreds to thousands of kelvin},
\newblock \bibinfo{journal}{The Journal of Physical Chemistry A}
  \bibinfo{volume}{115} (\bibinfo{year}{2011}) \bibinfo{pages}{6805--6812}.
\bibitem[{Fletcher et~al.(2018)Fletcher, Gustafsson, and
  Orton}]{fletcher_h2-h2}
\bibinfo{author}{L.~N. Fletcher}, \bibinfo{author}{M.~Gustafsson},
  \bibinfo{author}{G.~S. Orton},
\newblock \bibinfo{title}{Hydrogen dimers in giant-planet infrared spectra},
\newblock \bibinfo{journal}{The Astrophysical Journal Supplement Series}
  \bibinfo{volume}{235} (\bibinfo{year}{2018}) \bibinfo{pages}{24}.
\bibitem[{Abel et~al.(2012)Abel, Frommhold, Li, and Hunt}]{abel_h2-he}
\bibinfo{author}{M.~Abel}, \bibinfo{author}{L.~Frommhold},
  \bibinfo{author}{X.~Li}, \bibinfo{author}{K.~L. Hunt},
\newblock \bibinfo{title}{Infrared absorption by collisional h2--he complexes
  at temperatures up to 9000 k and frequencies from 0 to 20 000 cm$^{-1}$},
\newblock \bibinfo{journal}{The Journal of chemical physics}
  \bibinfo{volume}{136} (\bibinfo{year}{2012}) \bibinfo{pages}{044319}.
\bibitem[{Ptashnik et~al.(2019)Ptashnik, Solodov, and Solodov}]{PTASHNIK201997}
\bibinfo{author}{I.~V. Ptashnik}, \bibinfo{author}{A.~A. Solodov},
  \bibinfo{author}{A.~M. Solodov},
\newblock \bibinfo{title}{Fts measurements of the water vapour continuum
  absorption in 2.1$\mu$m atmospheric window},
\newblock \bibinfo{journal}{The XIX Symposium on High Resolution Molecular
  Spectroscopy 1–5 July}  (\bibinfo{year}{2019}). \URLprefix
  \url{https://symp.iao.ru/files/symp/hrms/19/presentation\_11485.pdf}.
\bibitem[{Ptashnik et~al.(2015)Ptashnik, Petrova, Ponomarev, Solodov, and
  Solodov}]{ptashnik2015b}
\bibinfo{author}{I.~V. Ptashnik}, \bibinfo{author}{T.~M. Petrova},
  \bibinfo{author}{Y.~N. Ponomarev}, \bibinfo{author}{A.~A. Solodov},
  \bibinfo{author}{A.~M. Solodov},
\newblock \bibinfo{title}{{Water vapor continuum absorption in near-IR
  atmospheric windows}},
\newblock \bibinfo{journal}{Atmospheric and Oceanic Optics}
  \bibinfo{volume}{28} (\bibinfo{year}{2015}) \bibinfo{pages}{115--120}.
  \DOIprefix\doi{10.1134/S1024856015020098}.
\bibitem[{Elsey et~al.(2020)Elsey, Coleman, Gardiner, Menang, and
  Shine}]{elsey20}
\bibinfo{author}{J.~Elsey}, \bibinfo{author}{M.~D. Coleman},
  \bibinfo{author}{T.~D. Gardiner}, \bibinfo{author}{K.~P. Menang},
  \bibinfo{author}{K.~P. Shine},
\newblock \bibinfo{title}{Atmospheric observations of the water vapour
  continuum in the near-infrared windows between 2500 and 6600 cm$^{-1}$},
\newblock \bibinfo{journal}{Atmos. Meas. Tech.} \bibinfo{volume}{13}
  (\bibinfo{year}{2020}) \bibinfo{pages}{2335--2361}.
  \DOIprefix\doi{10.5194/amt-13-2335-2020}.
\bibitem[{Ventrillard et~al.(2015)Ventrillard, Romanini, Mondelain, and
  Campargue}]{vent2015}
\bibinfo{author}{I.~Ventrillard}, \bibinfo{author}{D.~Romanini},
  \bibinfo{author}{D.~Mondelain}, \bibinfo{author}{A.~Campargue},
\newblock \bibinfo{title}{Accurate measurements and temperature dependence of
  the water vapor self-continuum absorption in the 2.1 $\mu$m atmospheric
  window},
\newblock \bibinfo{journal}{The Journal of Chemical Physics}
  \bibinfo{volume}{143} (\bibinfo{year}{2015}) \bibinfo{pages}{134304}.
  \DOIprefix\doi{10.1063/1.4931811}.
\bibitem[{Paynter et~al.(2009)Paynter, Ptashnik, Shine, Smith, McPheat, and
  Williams}]{Paynter09}
\bibinfo{author}{D.~J. Paynter}, \bibinfo{author}{I.~V. Ptashnik},
  \bibinfo{author}{K.~P. Shine}, \bibinfo{author}{K.~M. Smith},
  \bibinfo{author}{R.~McPheat}, \bibinfo{author}{R.~G. Williams},
\newblock \bibinfo{title}{{Laboratory measurements of the water vapor continuum
  in the 1200–8000 cm$^{-1}$ region between 293 K and 351 K}},
\newblock \bibinfo{journal}{Journal of Geophysical Research: Atmospheres}
  \bibinfo{volume}{114} (\bibinfo{year}{2009}) \bibinfo{pages}{D21301}.
  \DOIprefix\doi{https://doi.org/10.1029/2008JD011355}.
\bibitem[{Edwards et~al.(2019)Edwards, Mugnai, Tinetti, Pascale, and
  Sarkar}]{edwards_ariel}
\bibinfo{author}{B.~Edwards}, \bibinfo{author}{L.~Mugnai},
  \bibinfo{author}{G.~Tinetti}, \bibinfo{author}{E.~Pascale},
  \bibinfo{author}{S.~Sarkar},
\newblock \bibinfo{title}{{An Updated Study of Potential Targets for Ariel}},
\newblock \bibinfo{journal}{The Astronomical Journal} \bibinfo{volume}{157}
  (\bibinfo{year}{2019}) \bibinfo{pages}{242}.
  \DOIprefix\doi{10.3847/1538-3881/ab1cb9}.
\bibitem[{{Cowan} et~al.(2015){Cowan}, {Greene}, {Angerhausen}, {Batalha},
  {Clampin}, {Col{\'o}n}, {Crossfield}, {Fortney}, {Gaudi}, {Harrington},
  {Iro}, {Lillie}, {Linsky}, {Lopez-Morales}, {Mandell}, and
  {Stevenson}}]{cowan}
\bibinfo{author}{N.~B. {Cowan}}, \bibinfo{author}{T.~{Greene}},
  \bibinfo{author}{D.~{Angerhausen}}, \bibinfo{author}{N.~E. {Batalha}},
  \bibinfo{author}{M.~{Clampin}}, \bibinfo{author}{K.~{Col{\'o}n}},
  \bibinfo{author}{I.~J.~M. {Crossfield}}, \bibinfo{author}{J.~J. {Fortney}},
  \bibinfo{author}{B.~S. {Gaudi}}, \bibinfo{author}{J.~{Harrington}},
  \bibinfo{author}{N.~{Iro}}, \bibinfo{author}{C.~F. {Lillie}},
  \bibinfo{author}{J.~L. {Linsky}}, \bibinfo{author}{M.~{Lopez-Morales}},
  \bibinfo{author}{A.~M. {Mandell}}, \bibinfo{author}{K.~B. {Stevenson}},
\newblock \bibinfo{title}{{Characterizing Transiting Planet Atmospheres through
  2025}},
\newblock \bibinfo{journal}{Publications of the Astronomical Society of the
  Pacific} \bibinfo{volume}{127} (\bibinfo{year}{2015}) \bibinfo{pages}{311}.
  \DOIprefix\doi{10.1086/680855}.
\bibitem[{{Mugnai} et~al.(2020){Mugnai}, {Pascale}, {Edwards}, {Papageorgiou},
  and {Sarkar}}]{mugnai}
\bibinfo{author}{L.~V. {Mugnai}}, \bibinfo{author}{E.~{Pascale}},
  \bibinfo{author}{B.~{Edwards}}, \bibinfo{author}{A.~{Papageorgiou}},
  \bibinfo{author}{S.~{Sarkar}},
\newblock \bibinfo{title}{{ArielRad: the Ariel Radiometric Model}},
\newblock \bibinfo{journal}{Experimental Astronomy} \bibinfo{volume}{50}
  (\bibinfo{year}{2020}) \bibinfo{pages}{303--328}.
  \DOIprefix\doi{10.1007/s10686-020-09676-7}.
\bibitem[{Edwards and Stotesbury(2021)}]{Edwards_2021}
\bibinfo{author}{B.~Edwards}, \bibinfo{author}{I.~Stotesbury},
\newblock \bibinfo{title}{Terminus: A versatile simulator for space-based
  telescopes},
\newblock \bibinfo{journal}{The Astronomical Journal} \bibinfo{volume}{161}
  (\bibinfo{year}{2021}) \bibinfo{pages}{266}. \URLprefix
  \url{https://doi.org/10.3847/1538-3881/abdf4d}.
  \DOIprefix\doi{10.3847/1538-3881/abdf4d}.
\bibitem[{Gillon et~al.(2014)Gillon, Demory, Madhusudhan, Deming, Seager, Zsom,
  Knutson, Lanotte, Bonfils, Désert, and et~al.}]{Gillon_2014}
\bibinfo{author}{M.~Gillon}, \bibinfo{author}{B.-O. Demory},
  \bibinfo{author}{N.~Madhusudhan}, \bibinfo{author}{D.~Deming},
  \bibinfo{author}{S.~Seager}, \bibinfo{author}{A.~Zsom},
  \bibinfo{author}{H.~A. Knutson}, \bibinfo{author}{A.~A. Lanotte},
  \bibinfo{author}{X.~Bonfils}, \bibinfo{author}{J.-M. Désert},
  \bibinfo{author}{et~al.},
\newblock \bibinfo{title}{Search for a habitable terrestrial planet transiting
  the nearby red dwarf gj 1214},
\newblock \bibinfo{journal}{Astronomy \& Astrophysics} \bibinfo{volume}{563}
  (\bibinfo{year}{2014}) \bibinfo{pages}{A21}. \URLprefix
  \url{http://dx.doi.org/10.1051/0004-6361/201322362}.
  \DOIprefix\doi{10.1051/0004-6361/201322362}.
\bibitem[{Serdyukov et~al.(2016)Serdyukov, Sinitsa, and
  Lugovskoi}]{Serdyukov:16}
\bibinfo{author}{V.~I. Serdyukov}, \bibinfo{author}{L.~N. Sinitsa},
  \bibinfo{author}{A.~A. Lugovskoi},
\newblock \bibinfo{title}{Influence of gas humidity on the reflection
  coefficient of multilayer dielectric mirrors},
\newblock \bibinfo{journal}{Appl. Opt.} \bibinfo{volume}{55}
  (\bibinfo{year}{2016}) \bibinfo{pages}{4763--4768}. \URLprefix
  \url{http://ao.osa.org/abstract.cfm?URI=ao-55-17-4763}.
  \DOIprefix\doi{10.1364/AO.55.004763}.

\end{thebibliography}

\end{document}